\documentclass[letter,twocolumn]{jpsj2}
\def\runauthor{\textsc{S.~Watanabe} and \textsc{K.~Miyake}}
\makeatletter
\def\@typeset{}
\makeatother
\def\gsim{\mathop {\vtop {\ialign {##\crcr 
$\hfil \displaystyle {>}\hfil $\crcr \noalign {\kern1pt \nointerlineskip } 
$\,\sim$ \crcr \noalign {\kern1pt}}}}\limits}
\def\lsim{\mathop {\vtop {\ialign {##\crcr 
$\hfil \displaystyle {<}\hfil $\crcr \noalign {\kern1pt \nointerlineskip } 
$\,\,\sim$ \crcr \noalign {\kern1pt}}}}\limits}

\title{
Origin of Drastic Change of Fermi Surface and Transport Anomalies in 
CeRhIn$_5$ under Pressure 
}

\author{Shinji \textsc{Watanabe} 
and 
Kazumasa \textsc{Miyake}
}

\inst{
Division of Materials Physics, Department of Materials Engineering Science, Graduate School of
Engineering Science, Osaka University, Toyonaka, Osaka 560-8531, Japan
}

\abst{
The mechanism of 
drastic change of Fermi surfaces as well as transport anomalies 
near $P=P_{\rm c}\approx 2.35$~GPa in CeRhIn$_5$ is explained theoretically. 
The key mechanism is pointed out to be the interplay of magnetic 
order and Ce-valence fluctuations. 
We show that 
the antiferromagnetic state with ``small"  Fermi surfaces changes to 
the paramagnetic state with ``large" Fermi surfaces 
with huge enhancement of effective mass of electrons with keeping finite c-f hybridization. 
This explains the drastic change of the de Haas-van Alphen 
signals.  Furthermore, it is also consistent with the emergence of 
$T$-linear resistivity simultaneous with the residual resistivity peak 
at $P=P_{\rm c}$ in CeRhIn$_5$. 
}

\kword{CeRhIn$_5$, Fermi surface, non-Fermi liquid, localized-itinerant transition, valence fluctuation}

\begin{document}
\maketitle

The mechanism of instability of electronic states emerging 
when the magnetically-ordered temperature 
is suppressed to absolute zero by tuning material parameters 
has been one of the central issues in condensed matter physics~\cite{Moriya,Hertz,Millis}. 
A heavy-electron metal CeRhIn$_5$ has been extensively studied 
as the prototypical material. 
Accumulated experiments, however, 
have revealed that the physics of CeRhIn$_5$ seem to be
beyond conventional understanding~\cite{Moriya,Hertz,Millis} 
and require an essentially new concept.  

CeRhIn$_5$ undergoes an antiferromagnetic (AF) transition at $T_{\rm N}=3.8$~K 
with an ordered vector ${\bf Q}=(1/2, 1/2, 0.297)$ at ambient pressure~\cite{Bao}. 
When pressure is applied 
under a magnetic field larger than the upper critical field, the AF order is suppressed at 
$P=P_{\rm c}\approx 2.35$~GPa~\cite{Hegger,mito2001,Muramatsu,Knebel2004,Shishido2005,Park2006,Knebel2008}. 
Interestingly, 
a drastic change of Fermi surfaces was discovered 
at $P=P_{\rm c}$ by the de Haas-van Alphen (dHvA) measurement~\cite{Shishido2005}: 
For $0\le P< P_{\rm c}$, the main dHvA frequencies are 
in good agreement with those of LaRhIn$_5$ where the 4f electron is absent. 
On the other hand, for $P>P_{\rm c}$, the dHvA frequencies 
were identified to be approximately the same as those of CeCoIn$_5$, where 
4f electrons are itinerant, contributing to the formation of the Fermi surface. 
The electrons on a typical Fermi surface, the cylindrical $\beta_2$ blanch, 
shows the mass enhancement from $m^{*}=6m_0$ at $P=0$ to 60$m_0$ at $P\approx 2.2$~GPa 
with $m_0$ being a free-electron mass~\cite{Shishido2005}. 
For $P>P_{\rm c}$, the dHvA signal of the $\beta_2$ blanch is lost probably 
because of a too-large effective mass 
close to $100m_0$, where the heavy-electron state is realized 
in the paramagnetic-metal phase~\cite{Shishido2005}. 

So far, 
it has been thought that 
this drastic change of the Fermi surfaces might be explained in terms of the ``localized" to 
``itinerant" transition of f electrons~\cite{Park2008}. 
However, this conception encounters a serious difficulty in elucidating the experimental fact 
that the effective mass of electrons is enhanced even at $P=0$ 
with the Sommerfeld constant $\gamma\approx 56$~mJmol$^{-1}$K$^{-2}$~\cite{Hegger},
which is about 10 times larger than that of LaRhIn$_5$~\cite{Shishido2002,Philips2003}. 

Furthermore, resistivity measurements 
revealed that a striking anomaly emerges near $P=P_{\rm c}$~\cite{Muramatsu}: 
The low-$T$ resistivity $\rho(T=2.25~{\rm K})$ 
has a sharp peak at $P=P_{\rm c}$, 
suggesting that the residual resistivity $\rho_{0}$ is strongly enhanced 
near $P=P_{\rm c}$~\cite{Muramatsu,Knebel2008,Park2008}. 
The low-temperature resistivity also exhibits anomalous behavior: 
The $T$-linear registivity $\rho(T)\propto T$ emerges 
most prominently near $P=P_{\rm c}$ 
with a wide-$T$ range up to $\sim 10$~K~\cite{Muramatsu,Knebel2008,Park2008}. 
This behavior is quite different from that of normal metals described by the Fermi liquid $\rho\propto T^2$
as well as the conventional quantum criticality scenario 
near the AF quantum critical point (QCP) in three dimension (3D) $\rho\propto T^{1.5}$~\cite{Moriya,Hertz,Millis}. 
The origin and mechanism of 
these transport anomalies as well as the drastic change of the Fermi surface at $P=P_{\rm c}$ 
accompanied by huge mass enhancement have not been clarified so far.

Recently, an important measurement has been performed 
under pressure and magnetic field~\cite{Knebel2008}. 
G.~Knebel {\it et al}. have measured 
the $T^2$ coefficient $A$ in the resistivity in the $T\to 0$ limit 
at the magnetic field $H=15$~T, and 
found that $A$ increases as pressure increases toward $P=P_{\rm c}$. 
They have shown that under pressure, 
$\sqrt{A}$ shows an enhancement similar to the effective mass of electrons $m^{*}$ 
of the $\beta_2$ blanch obtained by the dHvA measurement~\cite{Shishido2005}, 
satisfying the $\sqrt{A}/m^{*}=$const. scaling. 
This indicates that the mass enhancement is {\it not} caused by the quantum criticality 
of the AF spin fluctuations, since 
in the case of the 3D (2D) AF QCP, 
$\sqrt{A}/m^{*}$ is expected to show the 
$T^{-1/4}$ $(T^{-1/2})$ divergence~\cite{Moriya,Hertz,Millis}. 
Namely, 
the $\sqrt{A}/m^{*}=$const. scaling strongly suggests that the mass enhancement near 
$P=P_{\rm c}$ purely comes from the band effect. 

On the basis of these observations, in this Letter, we present 
a theoretical explanation for resolving this outstanding puzzle in $\rm CeRhIn_5$. 
We show that the drastic change of the Fermi surface from a ``small" to a ``large" one 
occurs at the phase transition from AF to paramagnetic metal 
with huge mass enhancement under pressure, 
as observed. 
An important point here is that these results are obtained with finite hybridization 
between f and conduction electrons, which overcomes the difficulty of conventional scenario 
that f electrons undergo ``localized" to ``itinerant" transition~\cite{Park2008}. 
Our result not only naturally explains the $\sqrt{A}/m^{*}=$const. scaling, but also gives 
the reason why the $T$-linear resistivity emerges as well as the residual resistivity 
has a peak in the vicinity of $P=P_{\rm c}$. 

Let us start our discussion by introducing a minimal model, which describes 
the essential part of the physics of CeRhIn$_5$, in the standard notation: 
\begin{equation}
H=H_{\rm c}+H_{\rm f}+H_{\rm hyb}+H_{U_{\rm fc}}, 
\label{eq:PAM} 
\end{equation}
where 
$H_{\rm c}=\sum_{{\bf k}\sigma}\varepsilon_{\bf k}
c_{{\bf k}\sigma}^{\dagger}c_{{\bf k}\sigma}$ 
represents the conduction band, 
$H_{\rm f}=\varepsilon_{ \rm f}\sum_{i\sigma}n^{ \rm f}_{i\sigma}
+U\sum_{i=1}^{N}n_{i\uparrow}^{ \rm f}n_{i\downarrow}^{ \rm f}$ 
the f level and onsite Coulomb repulsion for f electrons, 
$H_{\rm hyb}=V\sum_{i\sigma}\left(
f_{i\sigma}^{\dagger}c_{i\sigma}+c_{i\sigma}^{\dagger}f_{i\sigma}
\right)$ 
the hybridization between f and conduction electrons, 
and 
$
H_{U_{\rm fc}}=
U_{\rm fc}\sum_{i=1}^{N}n_{i}^{ \rm f}n_{i}^{ c}
$ 
the Coulomb repulsion between f and conduction electrons, respectively. 
The $H_{U_{\rm fc}}$ term is a key parameter for explaining the anomalous transport 
properties of CeRhIn$_5$; 
The $\rho_0$ peak and $\rho\propto T$ observed in CeRhIn$_5$ 
are quite similar to the observations 
in CeCu$_2$Ge$_2$~\cite{jaccard}, CeCu$_2$Si$_2$~\cite{holms}, and 
CeCu$_2$(Si$_x$Ge$_{1-x}$)$_2$~\cite{yuan}. 
The pressure dependence of the coefficient 
$A$~\cite{jaccard,holms,yuan} as well as the Cu-NQR frequency~\cite{fujiwara} 
strongly suggest that these anomalies occur at the pressure where the valence of Ce changes sharply. 
The Ce-valence transition is well known as the $\gamma$-$\alpha$ transition in Ce metal. 
Band-structure calculation for Ce metal showed that 
4f- and 5d-electron bands are located at the Fermi level~\cite{Picket}. 
Since both the orbitals are located at the same Ce site, the inter-orbital Coulomb repulsion 
$U_{\rm fc}$ has a considerable magnitude, causing the first-order 
valence transition~\cite{WIM,watanabe2009}. 
Although $U_{\rm fc}$ is considered to be rather moderate in Ce compounds, 
valence fluctuations still affect physical quantities significantly even in such a 
valence-crossover regime~\cite{OM,M07,WTMF,watanabe2009}: 
Theoretical calculations based on the model~(\ref{eq:PAM}) 
have shown that strong Ce-valence fluctuations cause the 
enhancement of the residual resistivity~\cite{MM} as well as the $T$-linear resistivity~\cite{holms}. 

Since the $\rho_{0}$ peak and $\rho\propto T$ 
appear just at the boundary between the AF and paramagnetic phases, $P\approx P_{\rm c}$, in CeRhIn$_5$, 
the interplay of AF and the Ce-valence fluctuation seems to be a key mechanism~\cite{Knebel2004}. 
To treat both effects on equal footing, we apply the slave-boson mean-field theory~\cite{KR} to 
Eq.~(\ref{eq:PAM}). 
We use the hybridization form 
$VZ_{i\sigma}f_{i\sigma}^{\dagger}c_{i\sigma}$ 
instead of $Vf_{i\sigma}^{\dagger}c_{i\sigma}$ in Eq.~(\ref{eq:PAM}) 
by introducing bose creation (annihilation) operators $e_{i}^{\dagger} (e_i)$ and 
$d_{i}^{\dagger} (d_i)$ 
for the empty and doubly-occupied states, respectively, and 
$p_{i\uparrow}^{\dagger} (p_{i\uparrow})$ 
and $p_{i\downarrow}^{\dagger} (p_{i\downarrow})$ for singly-occupied states 
by requiring the constraint 
for completeness condition 
$
\sum_{i}\lambda_{i}'(e^{\dagger}_{i}e_{i}+p_{i\uparrow}^{\dagger}p_{i\uparrow}
+p_{i\downarrow}^{\dagger}p_{i\downarrow}+d_{i}^{\dagger}d_{i}-1)
$
and 
$
\sum_{i\sigma}\lambda_{i\sigma}
(f_{i\sigma}^{\dagger}f_{i\sigma}
-p_{i\sigma}^{\dagger}p_{i\sigma}-d_{i}^{\dagger}d_{i})
$ 
with $\lambda_{i}'$ and $\lambda_{i\sigma}$ being the Lagrange multipliers. 
Here, 
the renormalization factor for hybridization is defined as 
$Z_{i\sigma}\equiv(1-d_{i}^{\dagger}d_{i}-p_{i\sigma}^{\dagger}p_{i\sigma})^{-1/2}
(e_{i}^{\dagger}p_{i\sigma}+p_{i-\sigma}^{\dagger}d_i)
(1-e_{i}^{\dagger}e_{i}-p_{i-\sigma}^{\dagger}p_{i-\sigma})^{-1/2}$. 
To capture the essence of CeRhIn$_5$ as noted later, 
we consider the commensurate AF order on a bipartite lattice 
and consider $\lambda_{ia\sigma}$ and $\lambda_{ib\sigma}$ 
corresponding to the two sublattice ($ia$ and $ib$ sites) 
instead of $\lambda_{i\sigma}$, 
which are expressed as 
$\lambda_{ia\uparrow}=\lambda_{ib\downarrow}\equiv\lambda_i+\delta\lambda_i$, 
$\lambda_{ia\downarrow}=\lambda_{ib\uparrow}\equiv\lambda_i-\delta\lambda_i$, 
respectively. 
For $H_{U_{\rm fc}}$ in Eq.~(\ref{eq:PAM}), we employ the mean-field decoupling as 
$n^{\rm f}_{i}n^{\rm c}_{i}\simeq \bar{n}_{\rm f}n^{\rm c}_{i}+
\bar{n}_{\rm c}n^{\rm f}_{i}-\bar{n}_{\rm f}\bar{n}_{\rm c}$ 
with $\bar{n}_{\rm f}\equiv\sum_{i}\langle n_{i}^{\rm f} \rangle/N$ and 
$\bar{n}_{\rm c}\equiv\sum_{i}\langle n_{i}^{\rm c} \rangle/N$. 
By approximating mean fields and Lagrange multipliers as uniform ones, i.e., 
$e=\langle e_{i}\rangle$, $p_{\sigma}=\langle p_{i\sigma}\rangle$, and 
$d=\langle d_{i}\rangle$, $\lambda'=\lambda_i', \delta\lambda=\delta\lambda_i$, and 
$\lambda=\lambda_i$, respectively, 
the set of mean-field equations is obtained by 
optimal conditions 
$\partial\langle H \rangle/\partial x=0$ for 
$x=\lambda', \delta\lambda, \lambda, e, d, p_{\uparrow}$, and 
$p_{\downarrow}$:
\\
$e^2+p_{\uparrow}^2+p_{\downarrow}^2+d^2-1=0$, 
\\
$p_{\uparrow}^2-p_{\downarrow}^2=\frac{2}{N}
\sum_{{\bf k}}'\left[
 \langle f^{\dagger}_{{\bf k}\uparrow}f_{{\bf k}+{\bf Q}\uparrow} \rangle
-\langle f^{\dagger}_{{\bf k}\downarrow}f_{{\bf k}+{\bf Q}\downarrow} \rangle
\right]$
, 
\\
$p_{\uparrow}^2+p_{\downarrow}^2+2d^2=
\frac{1}{N}
\sum_{{\bf k}\sigma}'\left[
 \langle f^{\dagger}_{{\bf k}\sigma}f_{{\bf k}\sigma} \rangle
+\langle f^{\dagger}_{{\bf k}+{\bf Q}\sigma}f_{{\bf k}+{\bf Q}\sigma} \rangle
\right]$,
\\
$\frac{V}{N}
\sum_{{\bf k}\sigma}'\left(\frac{\partial Z_{\sigma}}{\partial e}
\right)
\left[
 \langle f^{\dagger}_{{\bf k}\sigma}c_{{\bf k}\sigma} \rangle
+\langle f^{\dagger}_{{\bf k}+{\bf Q}\sigma}c_{{\bf k}+{\bf Q}\sigma} \rangle
\right]+\lambda' e=0$, 
\\
$\frac{V}{N}
\sum_{{\bf k}\sigma}'\left(\frac{\partial Z_{\sigma}}{\partial d}
\right)
\left[
 \langle f^{\dagger}_{{\bf k}\sigma}c_{{\bf k}\sigma} \rangle
+\langle f^{\dagger}_{{\bf k}+{\bf Q}\sigma}c_{{\bf k}+{\bf Q}\sigma} \rangle
\right]
\\
\hspace*{4.7cm}
+(U+\lambda'-2\lambda) d=0$, 
\\
$\frac{V}{N}
\sum_{{\bf k}\sigma}'\left(\frac{\partial Z_{\sigma}}{\partial p_{\uparrow}}
\right)
\left[
 \langle f^{\dagger}_{{\bf k}\sigma}c_{{\bf k}\sigma} \rangle
+\langle f^{\dagger}_{{\bf k}+{\bf Q}\sigma}c_{{\bf k}+{\bf Q}\sigma} \rangle
\right]
\\
\hspace*{4.7cm}
+(\lambda'-\lambda-\delta\lambda)p_{\uparrow}=0$, 
\\
$\frac{V}{N}
\sum_{{\bf k}\sigma}'\left(\frac{\partial Z_{\sigma}}{\partial p_{\downarrow}}
\right)
\left[
 \langle f^{\dagger}_{{\bf k}\sigma}c_{{\bf k}\sigma} \rangle
+\langle f^{\dagger}_{{\bf k}+{\bf Q}\sigma}c_{{\bf k}+{\bf Q}\sigma} \rangle
\right]
\\
\hspace*{4.7cm}
+(\lambda'-\lambda+\delta\lambda)p_{\downarrow}=0$, 
\\
where ${\bf Q}$ is the AF-ordered vector. 
Here, $\sum_{\bf k}'$ denotes the summation over the 1st Brillouin zone. 
The chemical potential $\mu$ is determined so as to give the total filling 
$n=(\bar{n}_{\rm f}+\bar{n}_{\rm c})/2$ with 
$\bar{n}_{\rm f}+\bar{n}_{\rm c}=
\frac{1}{N}
\sum_{{\bf k}\sigma}'[
 \langle f^{\dagger}_{{\bf k}\sigma}f_{{\bf k}\sigma} \rangle
+\langle f^{\dagger}_{{\bf k}+{\bf Q}\sigma}f_{{\bf k}+{\bf Q}\sigma}\rangle
+\langle c^{\dagger}_{{\bf k}\sigma}c_{{\bf k}\sigma} \rangle
+\langle c^{\dagger}_{{\bf k}+{\bf Q}\sigma}c_{{\bf k}+{\bf Q}\sigma}\rangle
]$. 
These equations are solved self-consistently.

\begin{figure}[t]
\includegraphics[width=85mm]{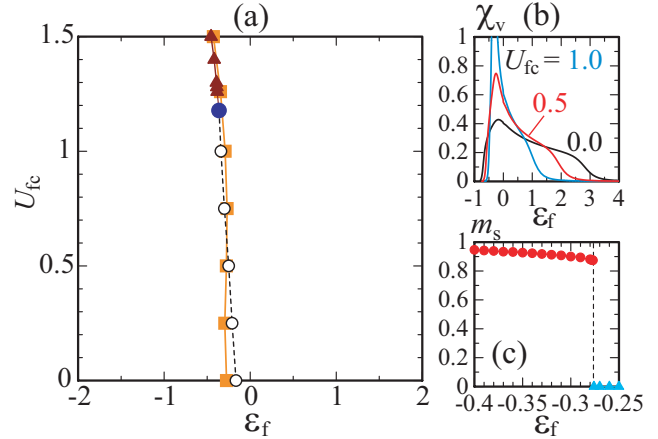}
\caption{\label{fig:PD}(color online) 
(a) Ground-state phase diagram in the plane of $U_{\rm fc}$ and $\varepsilon_{\rm f}$ 
for paramagnetic and AF states (see text). 
The first-order valence-transition line (solid line with triangles) 
terminates at the quantum critical point (filled circle). The valence crossover 
occurs at the dashed line with open circles, at which $\chi_{\rm v}$ has a maximum, 
as shown in (b). 
The solid line with filled squares represents the boundary between the AF state 
and the paramagnetic state (see text). 
(c) The AF order parameter $m_{\rm s}$ vs. $\varepsilon_{\rm f}$ 
for $U_{\rm fc}=0.5$. All results in (a)-(c) are calculated for 
$t=1$, $V=0.2$, and $U=\infty$ at $n=0.9$. 
}
\end{figure}

To clarify the mechanism of drastic change of the Fermi surface of CeRhIn$_5$, 
the most typical one, 
the two-dimensional-like Fermi surface, observed as the $\beta_2$ blanch, 
we consider $\varepsilon_{\bf k}=-2t(\cos k_x+\cos k_y)$ 
on the square lattice. Hereafter, the energy unit is taken as $t=1$. 
To simulate the AF order in the heavy-electron state realized in CeRhIn$_5$, 
we consider the small hybridization $V=0.2$ case near half filling $n=0.9$, 
and the AF order with ${\bf Q}=(\pi,\pi)$. 
Although we here calculated several $U$ cases, we show the result for $U=\infty$, 
since the result is essentially unchanged even for finite $U$, as far as $U$ is larger than the bandwidth.

First, we show the ground-state phase diagram determined under the assumption 
of the paramagnetic states with 
$p_{\uparrow}=p_{\downarrow}$ and $\delta\lambda=0$ in Fig.~\ref{fig:PD}(a). 
The first-order valence transition line (solid line with filled triangles) terminates 
at the QCP (filled circle) at 
$(\varepsilon_{\rm f}^{\rm QCP},U_{\rm fc}^{\rm QCP})=(-0.3623, 1.1778)$. 
For $U_{\rm fc}>U_{\rm fc}^{\rm QCP}$, $\bar{n}_{\rm f}$ shows a jump as a function of 
$\varepsilon_{\rm f}$, indicating the first-order transition between 
the paramagnetic metals with 
$\bar{n}_{\rm f}$ 
close to 1 and $\bar{n}_{\rm f}<1$ in deep-$\varepsilon_{\rm f}$ and 
shallow-$\varepsilon_{\rm f}$ regions, respectively, 
since large $U_{\rm fc}$ forces electrons to pour into either the f level or 
the conduction band~\cite{WIM,watanabe2009,OM,WTMF}. 
At the QCP, valence fluctuations diverge 
$\chi_{\rm v}=-\partial\bar{n}_{\rm f}/\partial\varepsilon_{\rm f}=\infty$, 
and for $U_{\rm fc}<U_{\rm fc}^{\rm QCP}$, $\chi_{\rm v}$ has a conspicuous peak 
at $\varepsilon_{\rm f}$ represented by the dashed line with open circles in Fig.~\ref{fig:PD}(a), 
indicating strong valence fluctuations, as shown in Fig.~\ref{fig:PD}(b). 
At the QCP, 
the characteristic energy scale of the system, the so-called Kondo temperature is given by 
$T_{\rm K}\equiv\bar{\varepsilon}_{\rm f}-\mu=2.93\times 10^{-3}$ with 
$\bar{\varepsilon}_{\rm f}=\varepsilon_{\rm f}+\lambda+U_{\rm fc}\bar{n}_{\rm c}$. 

When the AF states are taken into account, 
the AF order parameter defined as $m_{\rm s}\equiv p_{\uparrow}^2-p_{\downarrow}^2$ 
decreases as $\varepsilon_{\rm f}$ increases, 
as shown in Fig.~\ref{fig:PD}(c) for $U_{\rm fc}=0.5$. 
When the ground-state energies of this AF state and the paramagnetic state are 
comapared, the level crossing occurs at $\varepsilon_{\rm f}=\varepsilon_{\rm f}^{\rm c}$. 
Then, this AF and paramagnetic phase transition is identified to be of the first order. 
The phase boundary determined in this way is shown 
by the solid line with filled squares in Fig.~\ref{fig:PD}(a). 
We find that the AF order terminates in the vicinity of the first-order valence transition line 
and the valence-crossover line. 
These results 
imply that the suppression of the AF order occurs at the points with strong valence fluctuations.

These results are favorably compared with CeRhIn$_5$. 
Applying pressure to Ce systems corresponds to increasing $\varepsilon_{\rm f}$, since 
negative ions approach the tail of 4f wavefunction at the Ce site. 
Experimental fact of 
the sudden disappearance of the AF order at $P=P_{\rm c}$ and simultaneous 
emergence of the $\rho_{0}$ peak as well as $\rho\propto T$ near $P=P_{\rm c}$ 
seem to be well described by the results shown in Fig.~\ref{fig:PD}: 
The vicinity of $\varepsilon_{\rm f}^{\rm c}$
for moderate  $U_{\rm fc}(<U_{\rm fc}^{\rm QCP})$ with well-developed $\chi_{\rm v}$ 
seems to correspond to the vicinity of $P=P_{\rm c}$ in CeRhIn$_5$. 

\begin{figure}[t]
\includegraphics[width=88mm]{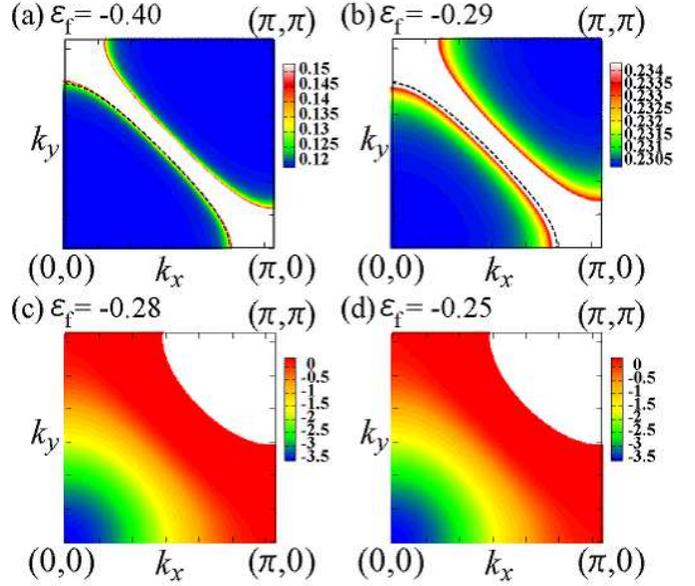}
\caption{\label{fig:FS}(color online) 
The contour plot of the energy band with $\downarrow$ spin 
located at the Fermi level $\mu$ 
for $t=1$, $V=0.2$, $U=\infty$, $U_{\rm fc}=0.5$, and $n=0.9$ at $h=0.005$: 
(a) $\varepsilon_{\rm f}=-0.40$, (b) $\varepsilon_{\rm f}=-0.29$, 
(c) $\varepsilon_{\rm f}=-0.28$, and (d) $\varepsilon_{\rm f}=-0.25$. 
The $E_{{\bf k}\downarrow}>\mu$ parts are represented by white regions. 
In (a) and (b), the dashed line indicates the Fermi surface of the conduction band, 
$\varepsilon_{\bf k}$ for $\bar{n}_{\rm c}=0.8$. 
}
\end{figure}

To analyze the Fermi-surface change shown by the dHvA measurement in CeRhIn$_5$, 
we apply the magnetic field to 
Eq.~(\ref{eq:PAM}) as $-h\sum_{i}(S_{i}^{{\rm f}z}+S_{i}^{{\rm c}z})$. 
The dHvA effect~\cite{Shishido2005} and $A$ coefficient~\cite{Knebel2008} 
were measured at $H=12\sim 17$~T and $H=15$~T, respectively. 
The magnetic field of $H=15$~T is estimated to be $h=0.0046t$ when 
the half bandwidth of the conduction band $4t$ of Eq.~(\ref{eq:PAM}) is compared with 
that of CeRhIn$_5$ by 
the band-structure calculation, about 1.5~eV~\cite{Hall}. 
Then, we show in Fig.~\ref{fig:FS} 
the contour plot of the energy band of Eq.~(\ref{eq:PAM}) with $\downarrow$ spin 
located at the Fermi level $\mu$ 
for $U_{\rm fc}=0.5$ at $h=0.005$. 

In the AF state, the lower hybridized band of Eq.~(\ref{eq:PAM}) is 
folded, giving rise to the hole region emerging at the magnetic zone boundary 
connecting ${\bf k}=(0,\pi)$ and $(\pi,0)$, as shown in Fig.~\ref{fig:FS}(a). 
Here, in order to make a comparison with the ``small" Fermi surface, which consists of only 
conduction electrons, we plot the Fermi surface of the conduction band $\varepsilon_{\bf k}$ 
at the filling $\bar{n}_{\rm c}=0.8$ as the dashed line in Fig.~\ref{fig:FS}(a). 
This Fermi surface corresponds to that when the hybridization between f and conduction electrons 
is switched off, $V=0$ in Eq.~(\ref{eq:PAM}); 
Since f electrons for $\bar{n}_{\rm f}=1$ are located at the localized f level for $V=0$, 
extra electrons in $n=0.9$, i.e., $\bar{n}_{\rm c}=0.8$ are in the conduction band. 
We see that the Fermi surface of the AF state for $V=0.2$ is nearly the same as the ``small" Fermi surface 
represented by the dashed line. 
This corresponds to the experimental fact that the dHvA signals of CeRhIn$_5$ are very similar to 
those of LaRhIn$_5$ in the AF-ordered phase for $P\le P_{\rm c}$~\cite{Shishido2002,Shishido2005}. 

The shape of the Fermi surface close to the ``small" Fermi surface remains until 
$\varepsilon_{\rm f}$ reaches 
the AF-paramagnetic boundary 
$\varepsilon_{\rm f}^{\rm c}=-0.283$, as 
shown in Fig.~\ref{fig:FS}(b), 
which corresponds to CeRhIn$_5$ at $P\lsim P_{\rm c}$. 
When $\varepsilon_{\rm f}$ exceeds $\varepsilon_{\rm f}^{\rm c}$, the Fermi surface 
drastically changes, as shown in Fig.~\ref{fig:FS}(c) for $\varepsilon_{\rm f}=-0.280$; 
The folding of the lower hybridized band disappears 
and the ``large" Fermi surface recovers, which is clearly different from the ``small" Fermi surface 
shown in Figs.~\ref{fig:FS}(a) and (b). 
This ``large" Fermi surface remains in the paramagnetic phase, as shown in Fig.~\ref{fig:FS}(d) 
for $\varepsilon_{\rm f}=-0.250$.

\begin{figure}[t]
\includegraphics[width=88mm]{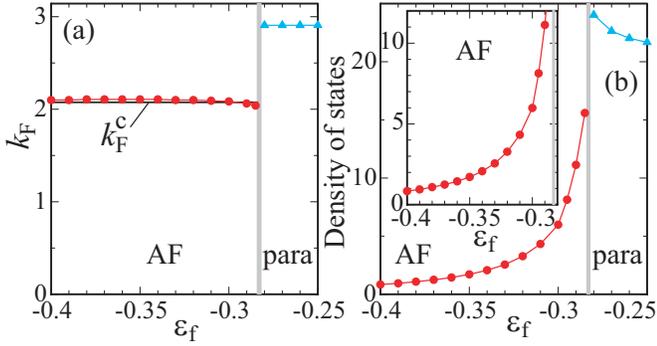}
\caption{\label{fig:DOS}(color online) 
(a) Fermi wavenumber $k_{\rm F}$ vs. $\varepsilon_{\rm f}$ in 
the AF state (circles) and the paramagnetic state (triangles) 
for $t=1$, $V=0.2$, $U=\infty$, $U_{\rm fc}=0.5$, and $h=0.005$ at $n=0.9$. 
The solid line represents $k_{\rm F}$ for the conduction band $\varepsilon_{\rm k}$ at 
$\bar{n}_{\rm c}=0.8$. 
(b) $D(\mu)$ vs. $\varepsilon_{\rm f}$ for the same parameters 
as (a). Inset is enlargement of the AF phase.
}
\end{figure}

To facilitate the comparison with the dHvA result, we plot the Fermi wave number $k_{\rm F}$, 
defined by the distance between ${\bf k}=(0,0)$ and the intersection point of the Fermi surface and 
the line connecting ${\bf k}=(0,0)$ and $(\pi,\pi)$, in Fig.~\ref{fig:DOS}(a). 
Here, for the AF-ordered state $\varepsilon_{\rm f}<\varepsilon_{\rm f}^{\rm c}$, $k_{\rm F}$ 
in the 1st Brillouin zone is plotted. 
We see that $k_{\rm F}$ is almost unchanged as a function of $\varepsilon_{\rm f}$, 
which is nearly the same as $k_{\rm F}$ of the conduction band at $\bar{n}_{\rm c}=0.8$, 
$k_{\rm F}^{\rm c}$ represented by the solid line. 
Then, $k_{\rm F}$ shows an abrupt jump at $\varepsilon_{\rm f}^{\rm c}$.
This is quite consistent with the dHvA measurement that the dHvA frequencies, including the $\beta_2$ blanch, 
keep almost constant for $0\le P<P_{\rm c}$, and they suddenly jump at $P=P_{\rm c}$ 
and remain constant for $P>P_{\rm c}$. 

We find that the measured enhancement of the effective mass of electrons 
from $m^{*}\sim 6m_0$ at $P=0$ to $m^{*}\sim 60m_0$ at $P\to P_{{\rm c}-}$~\cite{Shishido2005} 
is also reproduced; 
The $\varepsilon_{\rm f}$ dependence of 
the density of states (DOS) at the chemical potential $\mu$, $D(\mu)$, 
is shown in Fig.~\ref{fig:DOS}(b), where 
$D(\omega)\equiv\sum_{{\bf k}\sigma}\delta(\omega-E_{{\bf k}\sigma})/(2N)$ 
with $E_{{\bf k}\sigma}$ being the energy band of Eq.~(\ref{eq:PAM}). 
The DOS is enhanced about 10 times 
when $\varepsilon_{\rm f}$ approaches $\varepsilon_{\rm f}^{\rm c}$, 
as shown in the inset of Fig.~\ref{fig:DOS}(b). 
However, the renormalization factor $\sqrt{Z_{\sigma}}$, due to the many-body effect, does not show 
divergent growth even as $\varepsilon_{\rm f}$ approaches $\varepsilon_{\rm f}^{\rm c}$. 
This implies that the divergent  growth of the DOS is mainly due to the band effect. 
Then, $D(\mu)$ is proportional to $m^{*}$ and $\sqrt{A}$,  
explaining the measured $\sqrt{A}/m^{*}=$const. scaling~\cite{Knebel2008,MMV}. 
We note that $D(\mu)$ at $\varepsilon_{\rm f}=-0.4$ is about 10 times larger 
than the density of states of conduction electrons at $\bar{n}_{\rm c}=0.8$, $D_{\rm c}(\mu)=0.092$, 
which is also consistent with enhanced $\gamma$ of CeRhIn$_5$~\cite{Hegger} from 
that of 
LaRhIn$_5$~\cite{Shishido2002,Philips2003} at $P=0$.  
When $\varepsilon_{\rm f}$ increases, $m_{\rm s}$ decreases, as shown in Fig.~\ref{fig:PD}(c). 
Since the increase in $\varepsilon_{\rm f}$ 
tends to increase the renormalization factor $\sqrt{Z_{\sigma}}$, 
the energy gap between the original lower-hybridized band and the folded band in the AF phase 
is increased. 
This effect pushes up the latter band in the 1st Brillouin zone, 
making the flat part of the band, 
mainly contributed from f electrons, 
whose bottom is located at ${\bf k}=(0,0)$ start to emerge 
at the Fermi level (see Fig.~\ref{fig:FS}(b)). 

In the paramagnetic phase for $\varepsilon_{\rm f}>\varepsilon_{\rm f}^{\rm c}$, 
$D(\mu)$'s have larger values than those in the AF phase, 
as shown in Fig.~\ref{fig:DOS}(b). 
The increase in the DOS toward $\varepsilon_{\rm f}^{\rm c}$ in the paramagnetic phase 
is naturally understood 
since as $\varepsilon_{\rm f}$ decreases, 
$\bar{n}_{\rm f}$ increases to approach $1$, i.e., the Kondo state, 
giving rise to the reduction of $T_{\rm K}$, i.e., enhancement of $D(\mu)$. 
In CeRhIn$_5$, 
the dHvA signal of the $\beta_2$ blanch has not been detected for $P>P_{\rm c}$, 
probably because 
its effective mass is too large, close to $100m_{0}$~\cite{Shishido2005}.
This is also consistent with our result. 
For $\varepsilon_{\rm f}>\varepsilon_{\rm f}^{\rm c}$, 
$D(\mu)$ decreases as $h$ increases. 
This is also consistent with 
the field-dependence of $m^*$ of the $\beta_2$ blanch in CeCoIn$_5$~\cite{settai}, 
which is expected to correspond to the paramagnetic state of CeRhIn$_5$ for $P>P_{\rm c}$~\cite{M07,Pham}. 

The increase in $D(\mu)$ toward $\varepsilon_{\rm f}^{\rm c}$, shown in Fig.~\ref{fig:DOS}(b), 
suggests that total DOS at the 3D Fermi surface 
of CeRhIn$_5$ gives rise to the measured peak structure of the $A$ coefficient 
at $P=P_{\rm c}$~\cite{Knebel2008}. 
Our results clearly show 
that the ``small" Fermi surface can be observed by dHvA measurement even without 
switching off the c-f hybridization, which reminds us of the elucidation of metamagnetism 
in CeRu$_2$Si$_2$~\cite{wataKLM,Jullian,MI}. 

Recently, a possibility that the AF and paramagnetic transition is
continuous under pressure has been reported by the In-NQR measurement at
$H=0$~\cite{yashima2009}. To clarify the detailed nature of the phase
competition at the magnetic field smaller than the upper critical field, 
$\sim 10$~T, existence of the superconductivity 
should be taken into account, which is out of scope of the present study.

In summary, we have shown that the drastic change of Fermi surface with huge mass enhancement 
as well as the transport anomalies in CeRhIn$_5$ under pressure are naturally explained from 
the viewpoint of  the interplay of the AF order and Ce-valence fluctuations.



\end{document}